\documentclass[aps,prl,floatfix,nofootinbib,superscriptaddress,preprintnumbers,amsmath,amssymb,footinbib,twocolumn]{revtex4-1}
\usepackage{textcomp}
\usepackage{mathtools}
\usepackage{amsmath}
\usepackage{amssymb}
\usepackage{esint}
\usepackage{comment}
\usepackage[utf8]{inputenc} 

\usepackage[final]{graphics}
\usepackage{amsfonts}
\usepackage{color}
\usepackage{MnSymbol}

\renewcommand{\v}[1]{ \ensuremath{ {\mathbf{#1}_\perp} }}

\begin{document}

\title{Hierarchy of azimuthal anisotropy harmonics in collisions of small systems\\ from the Color Glass Condensate}

\author{Mark Mace}\affiliation{Physics Department, Brookhaven National Laboratory, Upton, New York 11973-5000, USA}\affiliation{Department of Physics and Astronomy, Stony Brook University, Stony Brook, NY 11794, USA}
\author{Vladimir V. Skokov}\affiliation{RIKEN-BNL Research Center, Brookhaven National Laboratory, Upton, New York 11973-5000, USA}
\author{Prithwish Tribedy}\affiliation{Physics Department, Brookhaven National Laboratory, Upton, New York 11973-5000, USA}
\author{Raju Venugopalan}\affiliation{Physics Department, Brookhaven National Laboratory, Upton, New York 11973-5000, USA}

\date{\today}
\begin{abstract}

We demonstrate that the striking systematics of two-particle azimuthal Fourier harmonics $v_2$ and $v_3$ in ultrarelativistic collisions of protons, deuterons and helium-3 ions off gold nuclei measured by the PHENIX Collaboration~\cite{Aidala:2018mcw} at the Relativistic Heavy Ion Collider (RHIC) is reproduced in the Color Glass Condensate (CGC) effective field theory. This contradicts the claim in~\cite{Aidala:2018mcw} that their data rules out initial state based explanations. The underlying systematics of the effect, as discussed previously in~\cite{Dusling:2017aot,Dusling:2017dqg,Dusling:2018hsg}, arise from the differing structure of strong color correlations between gluon domains of size $1/Q_S$ at fine ($p_\perp \gtrapprox Q_S$) or coarser ($p_\perp \lessapprox Q_S$) transverse momentum resolution. Further tests of the limits of validity of this framework can be carried out in light-heavy ion collisions at both RHIC and the Large Hadron Collider. Such measurements also offer novel opportunities for further exploration of the role of the surprisingly large short-range nuclear correlations measured at Jefferson Lab.

\end{abstract}

\maketitle

In a recent preprint~\cite{Aidala:2018mcw}, the PHENIX Collaboration presented measurements of the second and third Fourier harmonics ($v_2$ and $v_3$ respectively) of two-particle azimuthal correlations in collisions of protons ($p$), deuterons ($d$), and helium-3 ($^3$He) ions off gold (Au) nuclei at center-of-mass energies $\sqrt{s}=200$ GeV/nucleon. The measurements were performed in the 0-5\% centrality class of events in each of the three systems. They significantly improve the precision and reach of previous measurements~\cite{Aidala:2016vgl,Aidala:2017pup,Adare:2015ctn} and strongly confirm the system size dependence of the functions $v_2(p_\perp)$ and $v_3(p_\perp)$, where $p_\perp$ is the measured transverse momentum of charged hadrons. 

In \cite{Aidala:2018mcw}, the measurements are interpreted as providing strong support to the idea that the collisions of these small systems are producing nature's smallest droplets of a nearly perfect fluid Quark-Gluon Plasma (QGP)~\cite{Nagle:2018nvi}. This is in part due to the apparent agreement of the data with two hydrodynamical model computations SONIC~\cite{Habich:2014jna} and iEBE-VISHNU~\cite{Shen:2016zpp}. Further, while the transport model AMPT~\cite{Lin:2004en} reproduces their data, the authors of \cite{Aidala:2018mcw} suggest it is disfavored because model computations do not describe large and small systems with a consistent parameter set. Finally, \cite{Aidala:2018mcw} claims that initial state color correlations in the colliding ions are ruled out as an explanation of the systematics of their data. This claim is however not substantiated by any comparison to initial state models. 

In this Letter, we will demonstrate that initial state color correlations computed in the Color Glass Condensate~\cite{Gelis:2010nm} effective field theory (CGC EFT) describe the systematics of the PHENIX measurements of $v_{2,3}(p_\perp)$ in light-heavy ion collisions. The essential physics underlying the result was already noted in parton model computations wherein quarks from the small sized light ions scatter off domains of strong chromo-electromagnetic fields in the heavy ion target~\cite{Dusling:2017aot,Dusling:2017dqg,Dusling:2018hsg}. The size of the color domains in this ``toy" model are set by a semi-hard saturation scale $Q_S$ in the target. While this scale plays a critical role in what follows, many features  of the phenomena under consideration are universal and related to basic quantum properties of the underlying theory, including Bose-enhancement and Hanbury-Brown--Twiss (HBT) interference effects~\cite{Dumitru:2008wn,Kovchegov:2012nd,Altinoluk:2015uaa,Blok:2017pui,Kovchegov:2018jun,Kovner:2018vec,Altinoluk:2018hcu,Altinoluk:2018ogz}.

Our computations are performed within the dilute-dense power counting of the CGC EFT. 
Observables are computed in an expansion that includes the leading contribution and the first non-trivial saturation correction~\cite{Kovner:2016jfp,McLerran:2016snu,Kovchegov:2018jun} to the color charge density of the projectile and to all orders in the corresponding color charge densities of the dense Au nucleus. This saturation correction removes an ``accidental" parity symmetry arising from including only the leading order term, and is responsible for the $v_3$ azimuthal asymmetry in the dilute-dense approximation of the CGC EFT. The accidental nature of this symmetry was known previously from analytical and numerical computations in the full dense-dense (all orders in color charge densities of projectile and target) EFT~\cite{Lappi:2009xa,Schenke:2015aqa}. However because computations in the latter are numerically intensive, obtaining analytical expressions for the non-trivial saturation correction has proved extremely efficacious~\footnote{Comparing benchmark results in dense-dense and dilute-dense computations is a good measure of a systematic uncertainty in the latter.}.

\begin{figure*}[t]
\includegraphics[width=0.32\linewidth]{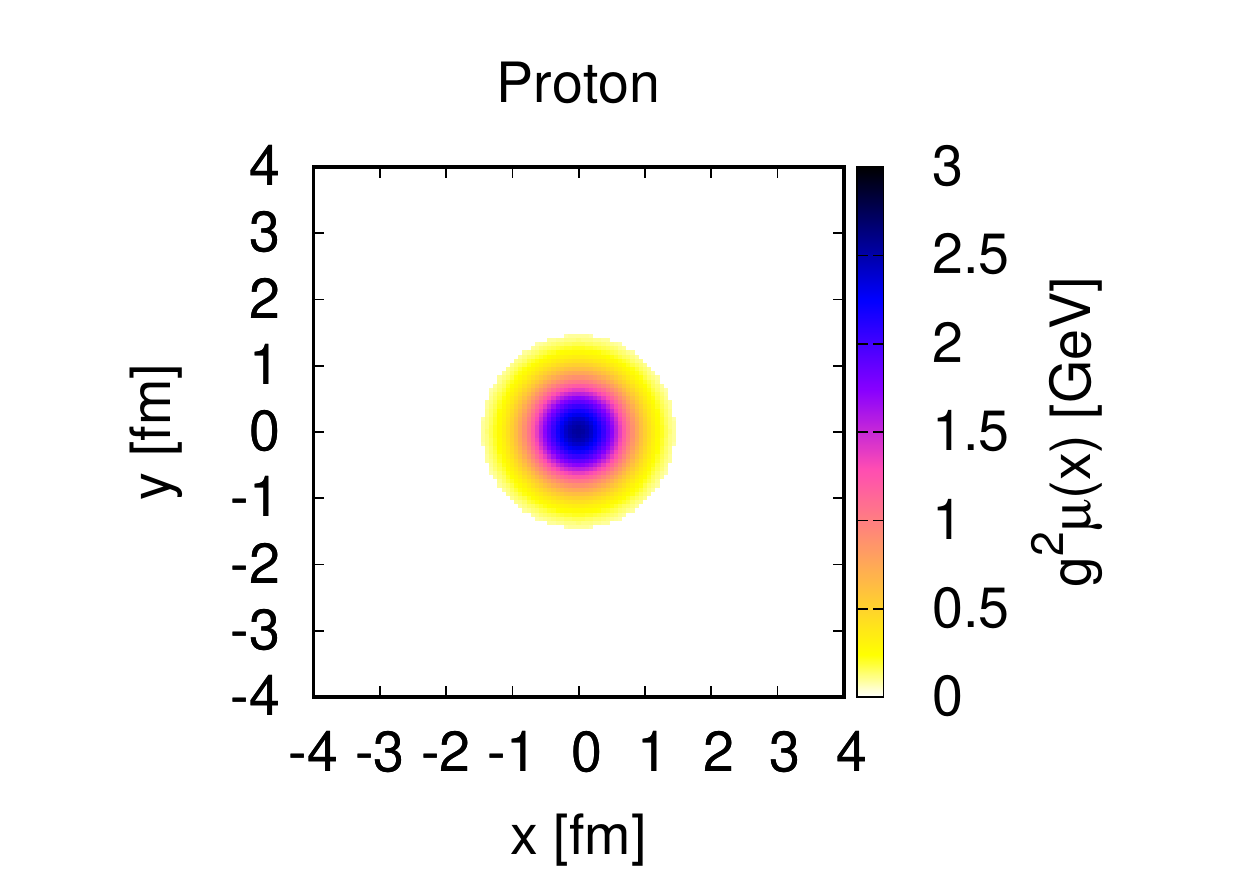}
\includegraphics[width=0.32\linewidth]{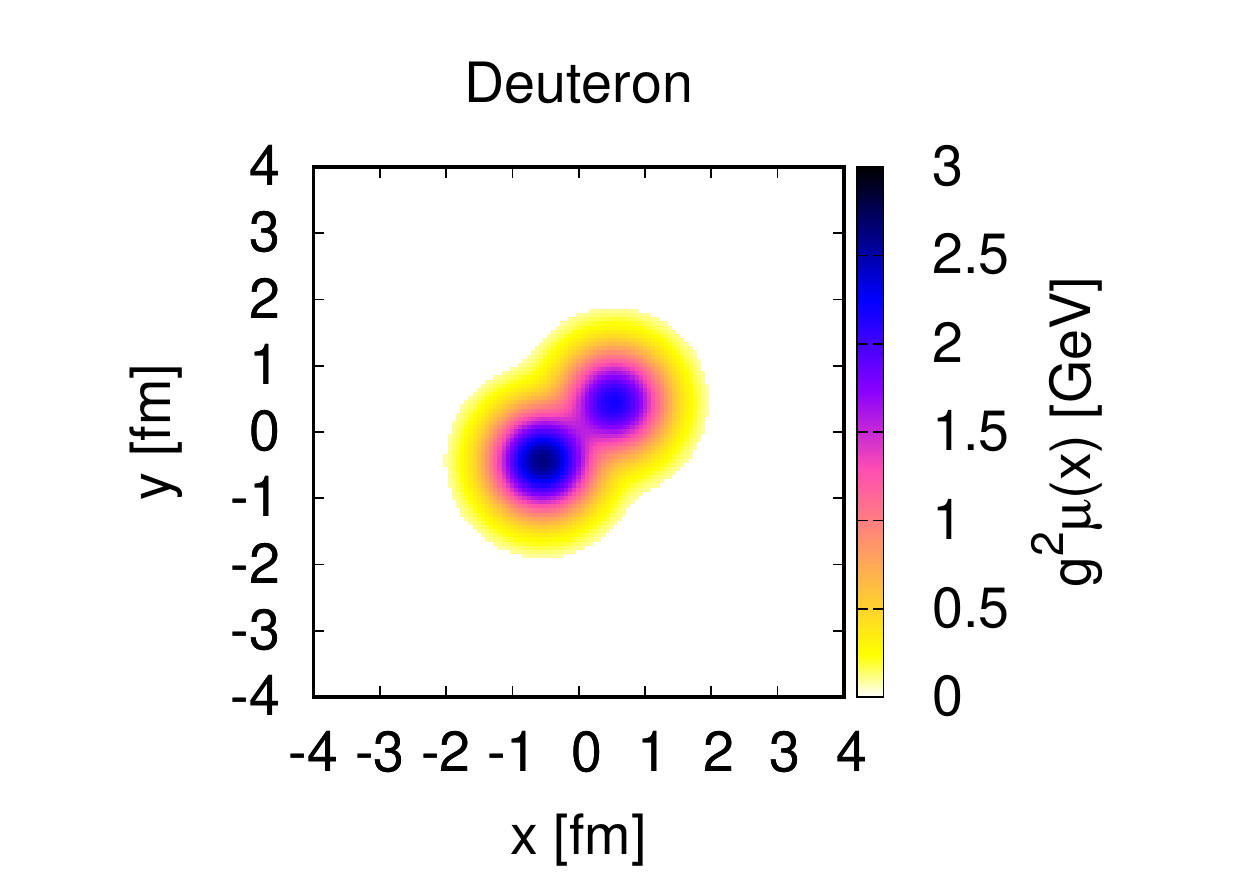}
\includegraphics[width=0.32\linewidth]{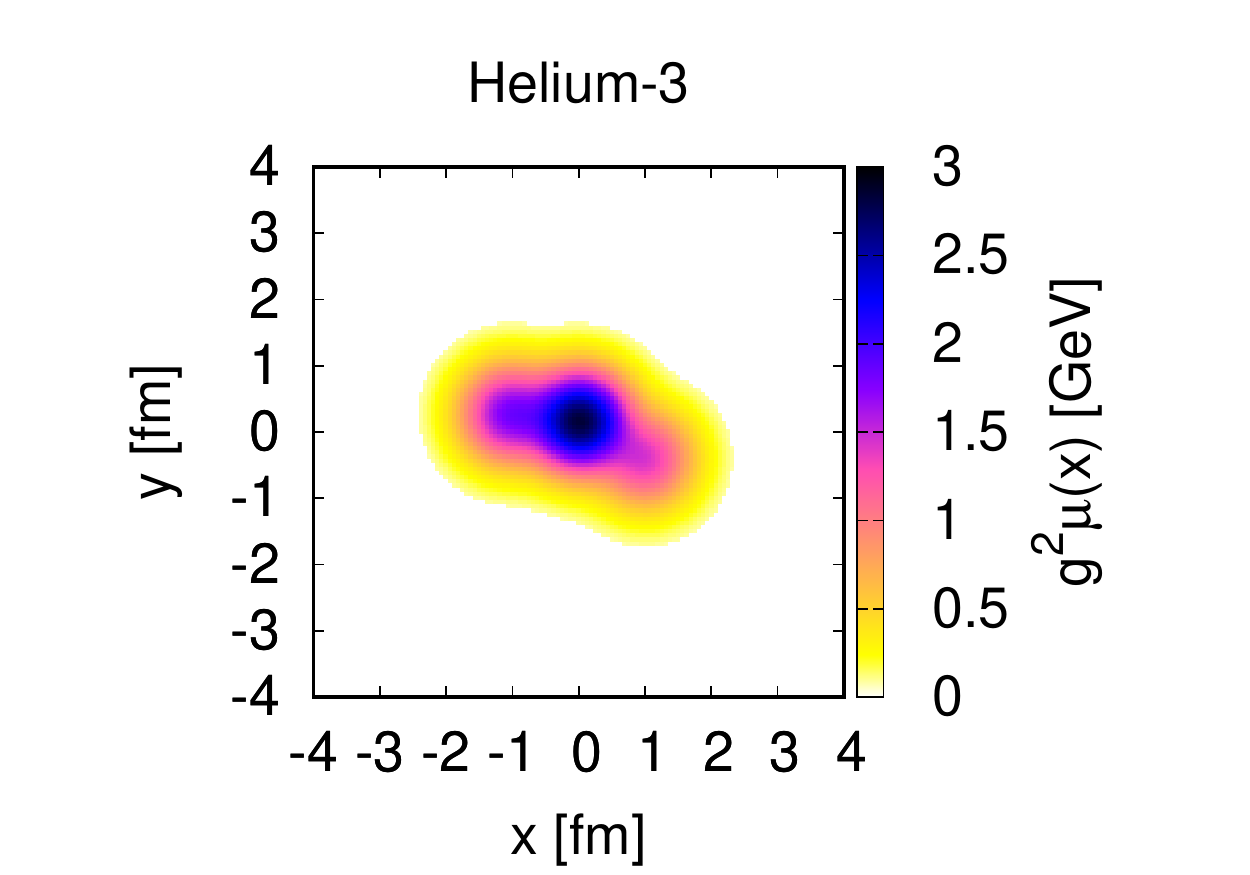}

\caption{ 
	Examples of color charge densities determined from Glauber sampling with the IP-Sat model~\cite{Kowalski:2003hm,Schenke:2012wb} for a single event for $p$, $d$, and $^3$He from high multiplicity events which contribute to the 0-5\% centrality class.
	\label{fig:ColorCharge}
}
\end{figure*}
The single particle inclusive gluon distribution in the dilute-dense CGC EFT, expressed as a functional of two-dimensional Fourier transform $\rho_p$ ($\rho_t$) of the projectile (target) color charge density, $\tilde\rho_p$ ($\tilde\rho_t$), can be generically decomposed into the parity-even and parity-odd contributions,
\begin{widetext}
\begin{align}
	%&\frac{dN (\v{k})}{d^2k dy}\Big[\rho_p, \rho_t\Big] = 
	%\frac{dN^{\rm even}(\v{k})}{d^2k dy}\Big[\rho_p, \rho_T\Big] +  
	%\frac{dN^{\rm odd}(\v{k})}{d^2k dy}\Big[\rho_p, \rho_T\Big]\,, \quad \quad %\notag\\ &
	\frac{dN^{\rm even,\, odd}(\v{k})}{d^2k dy}%\Big[\rho_p, \rho_t\Big] 
	=  
	\frac12 \left( \frac{dN (\v{k})}{d^2k dy}\Big[\rho_p, \rho_t\Big] \pm 
	\frac{dN (-\v{k})}{d^2k dy}\Big[\rho_p, \rho_t\Big] \right). 
\end{align}
Analytical computations~\cite{Kovchegov:1998bi,Dumitru:2001ux,Blaizot:2004wu,Kovner:2012jm,Kovchegov:2012nd} provide the compact result~\cite{McLerran:2016snu,Kovchegov:2018jun} 
\begin{align}
	\label{even}
	\frac{d N^{\rm even} (\v{k})}{d^2k dy} \Big[\rho_p, \rho_t\Big] &=  \frac{2}{(2\pi)^3} 
		\frac{ \delta_{ij} \delta_{lm}  +  \epsilon_{ij} \epsilon_{lm} }{k^2} 
		\Omega^a_{ij} (\v {k})
		\left[ \Omega^a_{lm} (\v {k}) \right]^\star\,,\\ 
	\label{odd}
	  \frac{d N^{\rm odd} (\v{k})}{d^2 k dy} \Big[\rho_p, \rho_t  \Big] 
    &=  
	{ \frac{2}{(2\pi)^3} }
	{\rm Im}
	\Bigg\{
		\frac{g}{{\v{ k}}^2} 
		\int \frac{d^2 l}{(2\pi)^2} 
				\frac{  {\rm Sign}({\v{k}\times \v{l}}) }{l^2 |\v{k}-\v{l}|^2 } 
		f^{abc}
			\Omega^a_{ij} (\v{l}) 
			\Omega^b_{mn} (\v{k}-\v{l})
			\left[\Omega^{c}_{rp} (\v{k})\right]^\star
		 \\ & \times \quad 
		\left[
			\left( 
			{\v{ k}}^2 \epsilon^{ij} \epsilon^{mn}
		-\v{l} \cdot (\v{k} - \v{l} ) 
		(\epsilon^{ij} \epsilon^{mn}+\delta^{ij} \delta^{mn}) 
		\right) \epsilon^{rp}+ 
		2 \v{k} \cdot (\v{k}-\v{l}) { \epsilon^{ij} \delta^{mn}} \delta^{rp}
		\right]
	\Bigg\} \, ,\notag 
\end{align}
\end{widetext}
where 
$
	\Omega_{ij}^a(\v{k}) = 
	g \int \frac{d^2 p }{(2\pi)^2}
	\frac{p_{i} (k-p)_{j} }{p^2}
	\rho^b_p(\v{p}) U_{ab} (\v{k}-\v{p}) 
$ and $\epsilon_{ij} (\delta_{ij})$ denotes the  Levi-Civita symbol (Kronecker delta). The adjoint Wilson line $U_{ab}$ is a functional of the target charge density and is the two-dimensional Fourier transform of its coordinate space counterpart:
$ \tilde U(\v{x}) = {\cal P} \exp \left( i g^2 \int dx^+ \frac{1}{\v{\nabla}^2}{\tilde \rho}_t^a(x^+, \v{x}) T_a \right). $

Comparing the even and odd contributions in Eqs.~\eqref{even} and \eqref{odd} respectively, one observes that the 
odd contribution is suppressed in the CGC EFT by $\alpha_S \rho_p$, where $\alpha_S = g^2/4\pi$ is the QCD coupling. This factor arises from the first saturation correction in the interactions with the dilute projectile ~\cite{McLerran:2016snu,Kovchegov:2018jun}. This systematic suppression in the power counting is what naturally explains in this framework the relative magnitude of $v^2_3\{2\}$ compared to $v^2_2\{2\}$ observed in the experimental data on small systems.  

\begin{widetext}
The $m$-particle momentum distribution is obtained after performing an ensemble average over the color charge distributions with the weight functionals, $W[\tilde\rho_{p,t}]$,
\begin{equation}
	\frac{ d^m N}{d^2k_1 dy_1 \cdots d^2k_m dy_m} = \int {\cal D} \rho_p {\cal D} \rho_t\  
	W[\rho_p] W[\rho_t] \  
	\frac{d N }{d^2k_1 dy_1} \Big[\rho_p, \rho_t\Big] 
	\cdots 
	\frac{d N }{d^2k_m dy_m} \Big[\rho_p, \rho_t\Big] \,.
\end{equation}
These have the form described by the McLerran-Venugopalan (MV) model~\cite{McLerran:1993ni,McLerran:1993ka}
\begin{equation}
	W[\tilde\rho_{p,t}] =  {\cal N} \exp\left[  -\int dx^{-,+} d^2 x \frac{\tilde\rho_{p,t}^a(x^{-,+} ,\v{x})  \tilde\rho_{p,t}^a  (x^{-,+} ,\v{x})  }{2 \mu^2_{p,t} }\right] \,,
\end{equation}
\end{widetext}
but are in fact more general because, as a consequence of renormalization group evolution of the color sources in the parton momentum fraction $x$~\cite{Dumitru:2011vk,Dusling:2009ni}, the color charge squared per unit area $\mu^2_{p,t}$ is a function of $x$ and spatial location of color charges in the transverse plane.  Specifically, we follow the same procedure as the phenomenologically constrained IP-Glasma model~\cite{Schenke:2012wb}, where the projectile and the target sources are placed using Glauber sampling of nucleons in the transverse plane~\cite{Loizides:2014vua}, with position $\mathbf{x}_\perp$ and impact parameter $\mathbf{b}_\perp$, and $g^2\mu(x,\mathbf{x}_\perp,\mathbf{b}_\perp)$ is determined by the IP-Sat model~\cite{Kowalski:2003hm}. Examples of the color charge distributions that produce gluons with multiplicities lying in the 0-5\% centrality class of the three systems are shown in Fig.~\ref{fig:ColorCharge}. We will return later to the important message conveyed by this visual depiction. 

An essential requirement of a first principles framework is to describe experimental data on the multiplicity distribution in light-heavy ion collisions. This is important to show that the framework captures the underlying physics of correlations and fluctuations. It is also crucial for performing a reliable ``apples to apples" centrality selection of events/configurations. It was shown in a remarkable paper~\cite{Gelis:2009wh}
that the CGC EFT generates negative binomial distributions (NBD);  subsequently, the impact parameter convoluted NBDs from the CGC EFT were employed to describe multiplicity distributions in proton-proton, proton-nucleus, and nucleus-nucleus collisions~\cite{Tribedy:2011aa,Schenke:2012fw}. Fluctuations of the saturation momentum itself~\cite{Iancu:2007st} are important to describe high multiplicity tails;  the quantitative impact of these is discussed in \cite{Schenke:2013dpa,McLerran:2015qxa}. 

%~~~~~~~~~~~~~~~~Figure~~~~~~~~~~~~~~~~~~~~~~~~~~~~~~~~~~~~~
\begin{figure}
\begin{center}
\includegraphics[width=0.95\linewidth]{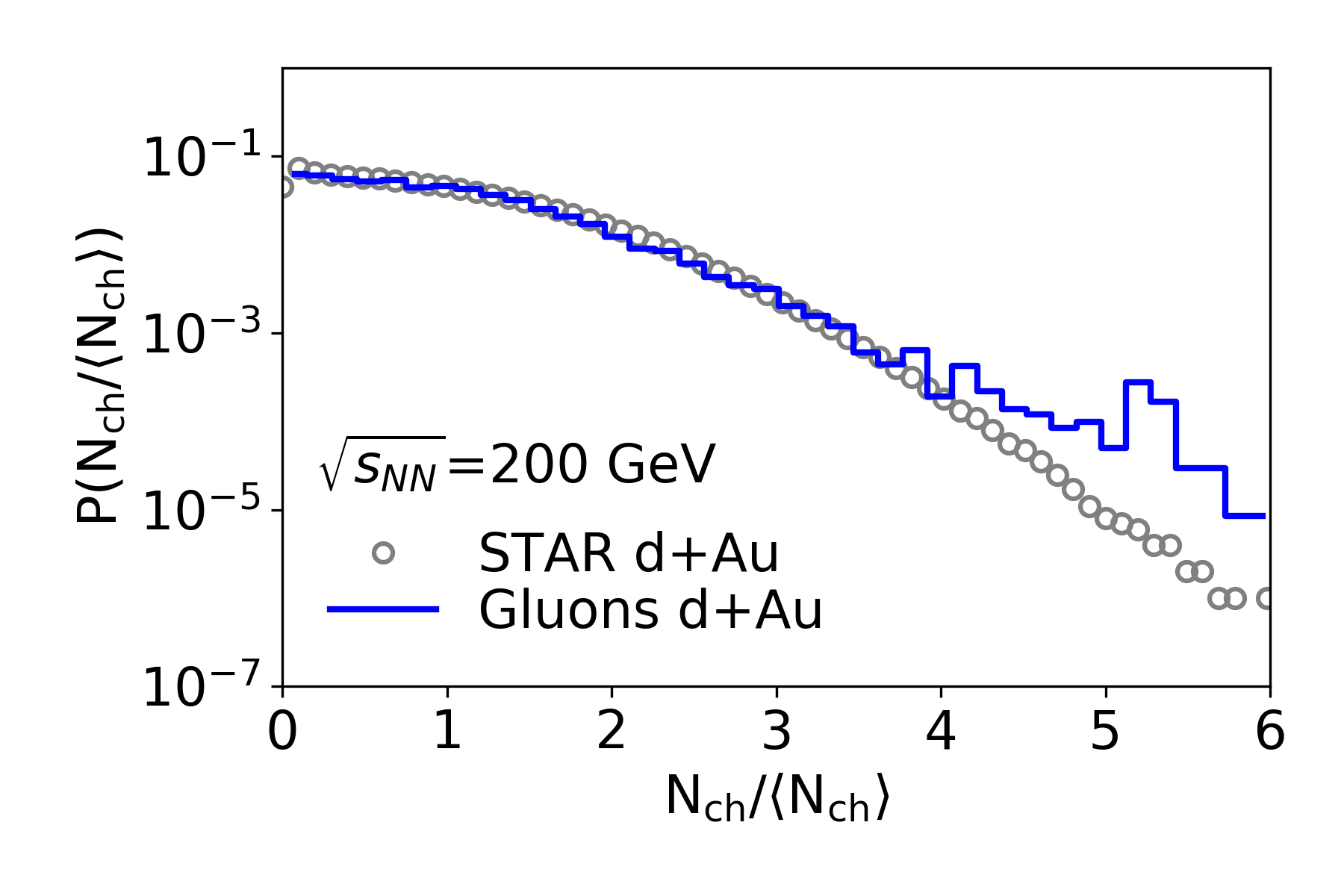}
\caption{ 
	The multiplicity distribution of produced particles computed in the dilute-dense CGC framework compared to STAR $d$+Au data~\cite{Abelev:2008ab}. 
	\label{fig:mult}
}
\end{center}
\end{figure}
%~~~~~~~~~~~~~~~~~~~~~~~~~~~~~~~~~~~~~~~~~~~~~~~~~~~~~~~~~~~

Figure~\ref{fig:mult} shows that the multiplicity distribution as a function of the number of charged hadrons $N_{\rm ch}$ for the rapidity window $|\eta|<0.5$ in $d$+Au collisions published by the STAR Collaboration at RHIC~\cite{Abelev:2008ab} is well reproduced in the dilute-dense CGC EFT. The details of the numerical computation on two-dimensional lattices are identical to those articulated previously~\cite{Lappi:2007ku,Schenke:2012wb,Dumitru:2014vka,Kovchegov:2018jun}. The free parameters in our framework are fixed by minimizing the deviations from the measured multiplicity distribution. These include the mean of the ratio $Q_S/g^2\mu$ taken to be $0.5$, the variance of  Gaussian fluctuations of $\text{ln}(Q_S^2)$~\cite{McLerran:2015qxa} taken to be $\sigma=0.5$, as well as an  infrared cutoff scale for color fields taken to be $m=0.3$ GeV. The effect of variations in these nonperturbative quantities was carefully examined in \cite{Schenke:2013dpa} and contributes to the systematic uncertainties of our computations.

With the parameters thus constrained, we now turn to computing the azimuthal anisotropies in light-heavy ion collisions. Defining for a fixed configuration of color sources, the harmonics for the  single particle azimuthal anisotropy as 
\begin{equation}
	V_n (p_1, p_2)  = \frac{ \int_{p_1}^{p_2}   k_\perp dk_\perp\frac{d\phi}{2\pi} e^{i n \phi} \frac{d N (\v{k})}{d^2k dy} \Big[\rho_p, \rho_t\Big] 
  }     {   \int_{p_1}^{p_2}   k_\perp dk_\perp \frac{d\phi}{2\pi}\frac{d N (\v{k})}{d^2k dy} \Big[\rho_p, \rho_t\Big] } \,,
\end{equation}
the physical two-particle anisotropy coefficients can be simply expressed as 
\begin{align}
	v_n^2\{2\} (p_\perp) 
	&=   \int {\cal D} \rho_p {\cal D} \rho_t\  W[\rho_p] \ W[\rho_t] \ 	
	\\ \notag 
	&\quad \times V_n (p_\perp-\Delta/2,p_\perp+\Delta/2) 
	V^\star_n (0, \Lambda_{\rm UV}) 
	\,.
\label{fig:vn-formula}
\end{align}
We consider $\Delta=0.5$ GeV bins in $p_\perp$, similar to what was done in~\cite{Aidala:2018mcw}. It is important to note that technically~\cite{Aidala:2018mcw} calculates the event plane $v_2$; however since the event plane resolution is small~\footnote{The PHENIX experiment has event plane resolution of 6.7\% and 5.7\% for $p$+Au and $d$+Au respectively~\cite{Aidala:2018mcw} }, these two quantities are very similar~\cite{Luzum:2012da}. Here $\Lambda_{\rm UV}$ is the ultraviolet $p_\perp$ cutoff, defined by the inverse lattice spacing -- our results are insensitive to this cutoff.
An identical computation was performed previously in the dense-dense framework to extract $v_2$ and $v_3$~\cite{Schenke:2015aqa}. Our dilute-dense framework, however, has the significant advantage that analytical expressions can be written down and results do not require numerical evaluation of the temporal evolution of the classical Yang-Mills equations. 

%~~~~~~~~~~~~~~~~Figure~~~~~~~~~~~~~~~~~~~~~~~~~~~~~~~~~~~~~
\begin{figure*}[]
\includegraphics[width=0.425\linewidth]{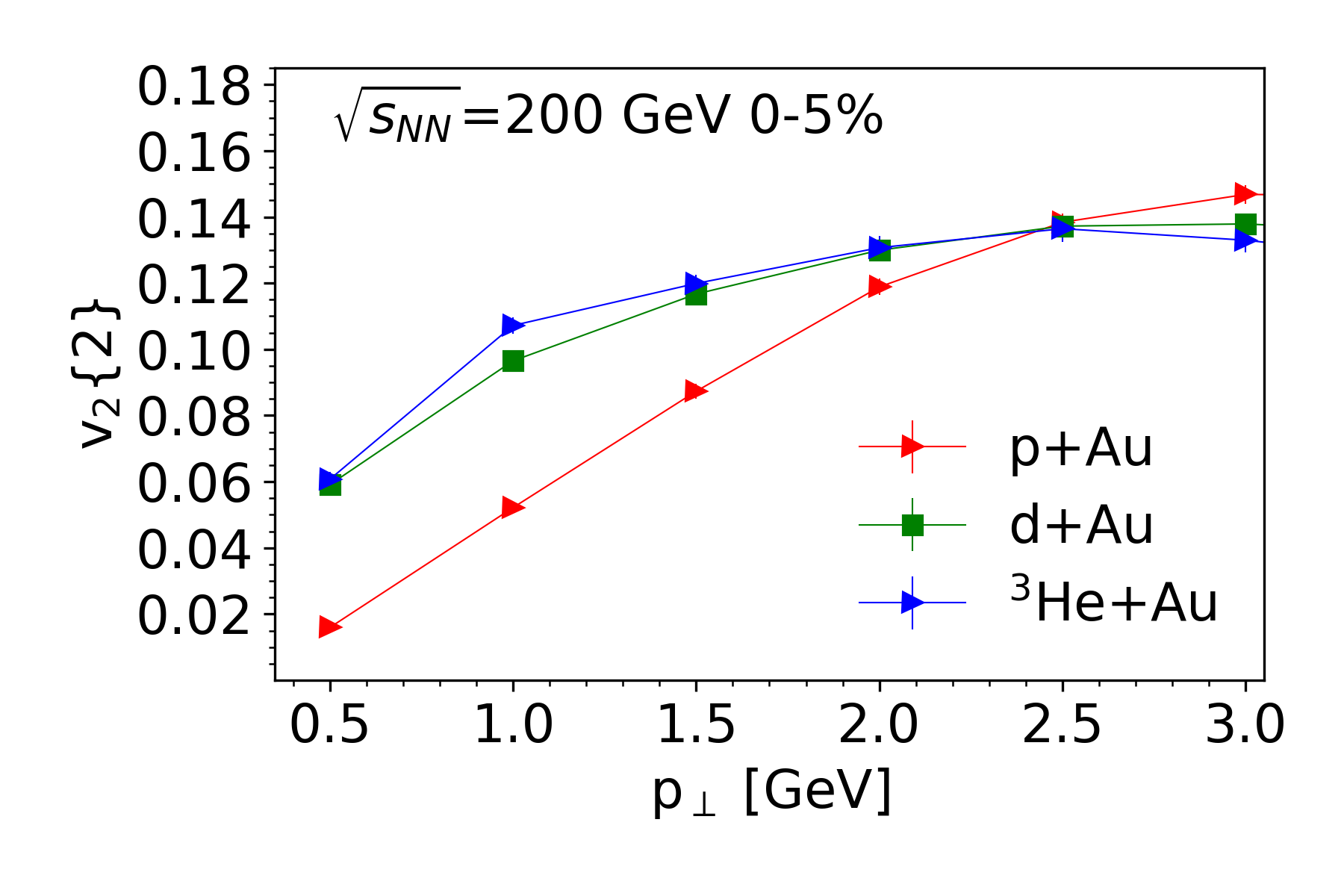}
\includegraphics[width=0.425\linewidth]{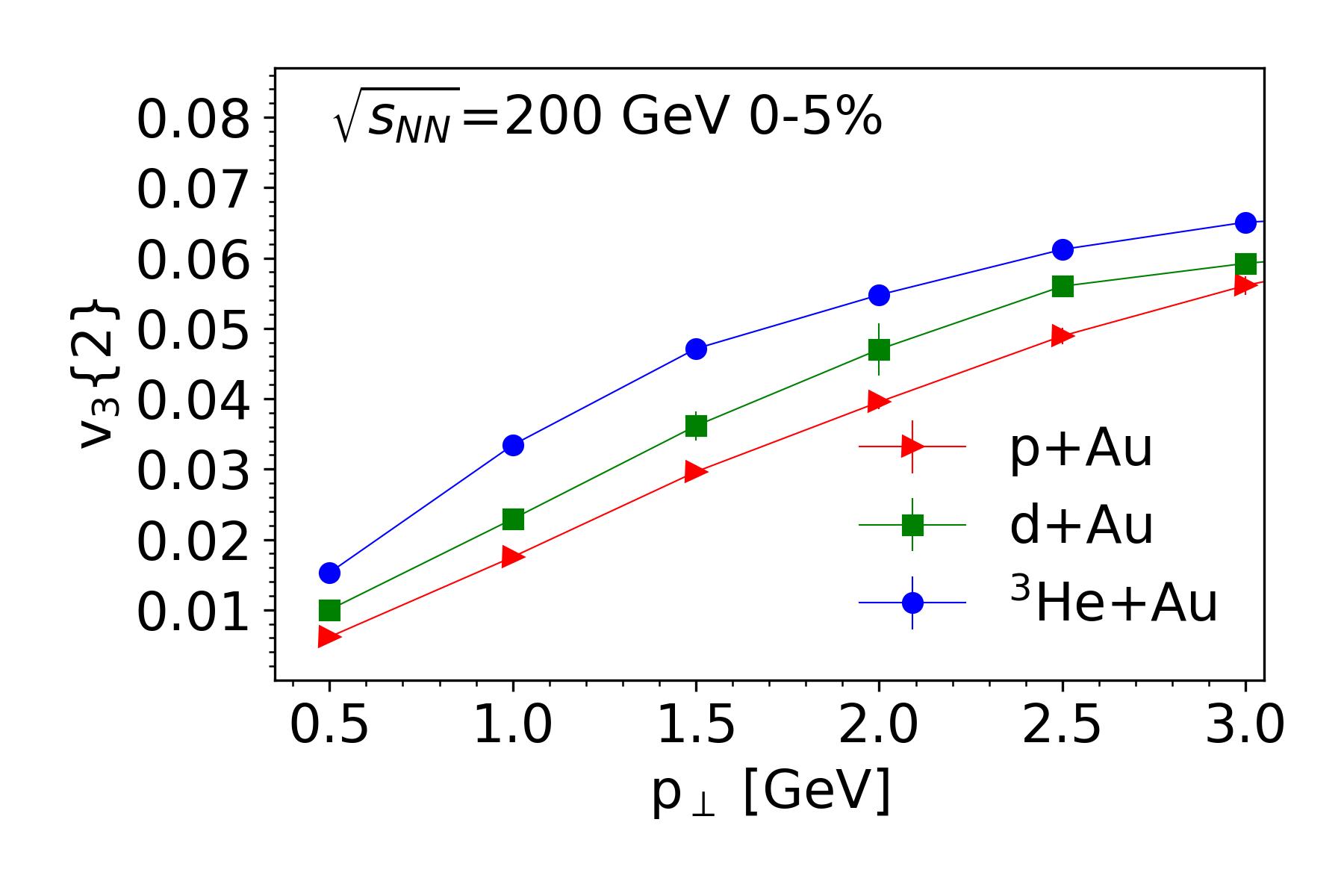}
\caption{ 
	Hierarchy  of  anisotropies $v_{2,3}(p_\perp)$ of gluons produced in the 0-5\% centrality class of light-heavy ion collisions computed in the 
	dilute-dense CGC framework 
	\label{fig:Hierarchy}
}
\end{figure*}
%~~~~~~~~~~~~~~~~~~~~~~~~~~~~~~~~~~~~~~~~~~~~~~~~~~~~~~~~~~~

The results of our computation for the hierarchy of $v_2(p_\perp)$ and $v_3(p_\perp)$ gluon anisotropies for $p$+Au, $d$+Au and $^3$He+Au collisions at $\sqrt{s}=200$ GeV/nucleon in the 0-5\% centrality class for each of the three systems is shown in Fig.~\ref{fig:Hierarchy}. A clear hierarchy is observed in the magnitudes of $v_{2,3}(p_\perp)$ for $p$+Au and those for $d$+Au and $^3$He+Au. For most of the $p_\perp$ range plotted, this is opposite to the naive expectation that the anisotropies should be suppressed with an increasing number of color domains. Between $d$+Au and $^3$He+Au, this hierarchy is not clearly distinguishable for $v_2(p_\perp)$; it is, however, clearly visible for $v_3(p_\perp)$. As discussed previously in the toy model computation of~\cite{Dusling:2017dqg,Dusling:2017aot,Dusling:2018hsg}, the scaling with inverse number of color domains is violated primarily because of the interplay of two dimensionful scales: in our case, these are $Q_S^p$, the saturation scale in the projectile, and a typical gluon resolution scale $\tilde{p}_\perp$. In the toy model of~\cite{Dusling:2017dqg,Dusling:2017aot,Dusling:2018hsg}, $\tilde{p}_\perp$ is identical to ${p}_\perp$; in the dilute-dense framework, their relation is less straightforward. Domain scaling holds if $\tilde{p}_\perp > Q_S^p$, because then gluons in the target resolve individual color domains in the projectile. This is clearly seen for $v_2(p_\perp)$ at high $p_\perp$ in Fig.~\ref{fig:Hierarchy} and a similar trend is seen for $v_3(p_\perp)$.  In contrast, if $\tilde{p}_\perp < Q_S^p$, gluons in the target cannot resolve individual domains any more but interact with $\sim (Q_S^p)^2/\tilde{p}_\perp^2$ of them simultaneously. In \cite{Dusling:2017aot}, it was shown that in this case there is no suppression with the number of color domains! 

Another important element in understanding the systematics of the data is that 0-5\% centrality in $^3$He+Au collisions corresponds to a significantly higher value of $N_{\rm ch}$ than for $p$+Au collisions. In the dilute-dense framework, the multiplicity of an event scales with $(Q_S^p)^2$~\cite{Dumitru:2001ux,Krasnitz:2002mn}. Thus for 0-5\% centralities, $Q_S^p|_{^3{\rm He}} > Q_S^p |_{\rm p}$. Hence, as long as $\tilde{p}_\perp < Q_S^p |_{^3{\rm He}}$, gluons in the target will coherently interact with $(Q_S^p|_{^3{\rm He}} )^2/\tilde{p}_\perp^2$ domains, many more than in $p$+Au. As shown in simple color domain models~\cite{Dumitru:2014yza,Lappi:2015vta}, the corresponding chromoelectric fields will generate larger anisotropies, but as the similar values of $v_2(p_\perp)$ between $d$+Au and $^3$He+Au in Fig.~\ref{fig:Hierarchy} indicates, $v_2(p_\perp)$ will saturate at large $N_{\rm ch}$. Because $v_3$ is due to a higher order $\alpha_S\rho_p$ suppressed effect, this saturation may only occur for larger $N_{\rm ch}$. Our prediction would therefore be that $v_{2,3}(p_\perp)$ for high multiplicity events across small systems should be identical for the same $N_{\rm ch}$.

%
%~~~~~~~~~~~~~~~~Figure~~~~~~~~~~~~~~~~~~~~~~~~~~~~~~~~~~~~~~
\begin{figure*}[]
\includegraphics[width=0.95\linewidth]{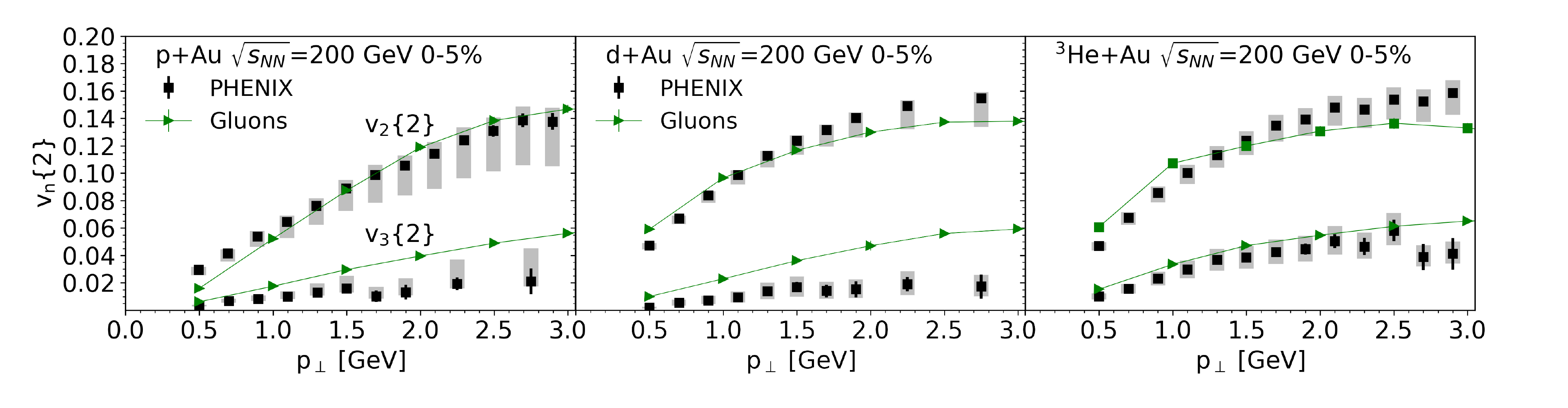}
\caption{ 
	Comparision of the results shown in Fig.~\ref{fig:Hierarchy} to $v_2$ and $v_3$ for charged hadron data from the PHENIX Collaboration~\cite{Aidala:2018mcw}.  
	\label{fig:phenix}
}
\end{figure*}
%~~~~~~~~~~~~~~~~~~~~~~~~~~~~~~~~~~~~~~~~~~~~~~~~~~~~~~~~~~~

In Fig.~\ref{fig:phenix}, we overlay Fig.~\ref{fig:Hierarchy} on top of the data for charged hadrons presented by the PHENIX Collaboration in ~\cite{Aidala:2018mcw}. The agreement for $v_2(p_\perp)$ is quite good across systems. 
This is interesting as $v_2(p_\perp)$ in hydrodynamical models is particularly sensitive to spatial geometry.
For $v_3(p_\perp)$, while the agreement for $^3$He is quite good, the computation overshoots the data in the $p$, $d$ systems. Since $v_3$ is fluctuation driven, we speculate this may be related to the fact that our comparison of gluon multiplicities to the $N_{\rm ch}$ multiplicity distribution in Fig.~\ref{fig:mult} also overshoots the data at high multiplicities. A corollary of this statement is that our $P(N_{ch}/\langle N_{ch} \rangle)$ for $^3{\rm He}$ should agree with the RHIC $N_{ch}$ distribution when available. Nevertheless, our computation and the data share the feature that $v_3(p_\perp)$ for $p$, $d$ is lower than that for $^3$He. We note that hydrodynamical models show a similar hierarchy for $v_3(p_\perp)$~\cite{Shen:2016zpp}. We emphasize again that a stronger prediction in our framework is that $v_3$ in high multiplicity small systems will agree for the same $N_{\rm ch}$. This is indeed what is seen in peripheral A+A collisions at the LHC and in central $p+A$ collisions at the same $N_{\rm ch}$~\cite{Chatrchyan:2013nka}, a feature of data which in that case is clearly hard to explain by system geometry alone~\cite{Basar:2013hea}. 

There are significant systematic uncertainties in the computation. Firstly, within the framework itself, there are higher order color charge density corrections, which may contribute differently for each of the projectile ions. This uncertainty can be benchmarked with numerical dense-dense computations for each species. Secondly, nonperturbative model parameters that are fixed for $d$+Au collisions by the measured multiplicity distributions may differ for multiplicity distributions of the other light ions. These are not available at present. There are uncertainties due to gluon fragmentation and higher order QCD computations. The technology to estimate the former exists within the CGC+PYTHIA framework~\cite{Schenke:2016lrs}, whereby gluons produced from the CGC are connected via strings and the latter are fragmented into hadrons with the PYTHIA event generator~\cite{Sjostrand:1984ic,Sjostrand:2014zea}. Finally, the relative contribution of final-state scattering should be quantified by matching the CGC initial state to transport models at later times~\cite{Greif:2017bnr}. 

While comparisons to data that progressively reduce the stated theoretical uncertainties are essential to understand the quantitative role of initial state correlations, qualitative trends in data may suffice to assess their dominant role.  As our discussion of the physical underpinnings of the anisotropies suggests, CGC EFT computations will generate simple systematics of $v_{2,3}\{2\}$, as a function of $N_{\rm ch}$ and $\sqrt{s}$ which should be straightforward to rule out. Further, multiparticle azimuthal anisotropy correlations $v_n\{m\}$ for $m\geq 4$ can be computed for light-heavy ion collisions and their systematics compared to data; we caution, however, that universal features of the mathematical structure of CGC EFT n-body distributions and those of hydrodynamic single particle distributions may lead to similar results~\cite{Yan:2013laa}. 

A key uncertainty is our knowledge of rare nuclear contributions in high multiplicity events. As suggested by Fig.~\ref{fig:ColorCharge}, nucleons overlap more closely in such events relative to minimum bias events. In particular, electron scattering experiments at Jefferson Lab have revealed that short-range pairing of nucleons dominates nuclear wavefunctions for momenta larger than the Fermi momentum~\cite{Hen:2016kwk,Cruz-Torres:2017sjy}. It is conceivable therefore that such ``clumpy" nucleon configurations may contribute significantly to the $^3$He nuclear wavefunction, beyond those anticipated from Green's function Monte Carlo computations~\cite{Carlson:1997qn,Nagle:2013lja}. This interesting possibility is under active investigation~\cite{OH-MM-AS-RV-in-proc}. 

In summary, we have shown that initial state color correlations in the dilute-dense framework of the CGC EFT provides a competitive explanation for the 
data presented on azimuthal anisotropy coefficients $v_2$ and $v_3$ in small collision systems. Further comparisons to data in different centrality classes, and to multiparticle anisotropies, can help quantify if and where the dominant role of initial state correlations in describing the collectivity observed in small systems breaks down. 

We thank Ron Belmont, Adam Bzdak, Adrian Dumitru, Kevin Dusling, Or Hen, Jiangyong Jia,  Alex Kovner, Yuri Kovchegov, Larry McLerran, Wei Li, Jean-Yves Ollitrault, Bj\"{o}rn Schenke, Chun Shen, Mike Strickland and Li Yi for useful discussions. We thank Sylvia Morrow for providing us the PHENIX data presented in~\cite{Aidala:2018mcw}. R. V. would in particular like to thank Juergen Schukraft for numerous illuminating conversations on the topic and for his gracious public acknowledgement of the results of a wager. This material is based on work supported by the U.S. Department of Energy, Office of Science, Office of Nuclear Physics, under Contracts No. DE-SC0012704 (M.M.,P.T.,R.V.) and DE-FG02-88ER40388 (M.M.). M.M. would also like to thank the BEST Collaboration for support. This research used resources of the National Energy Research Scientific Computing Center, a DOE Office of Science User Facility supported by the Office of Science of the U.S. Department of Energy under Contract No. DE-AC02-05CH11231.

%\bibliography{bibl_insp.bib}

\end{document}